\documentclass{emulateapj}

\usepackage{natbib}
\begin{document}
\title{Soft coronal X-rays from $\beta$ Pictoris}

\author{H.~M.~G\"unther, S.~J.~Wolk, J.~J.~Drake}
\affil{Harvard-Smithsonian Center for Astrophysics, 60 Garden Street, Cambridge, MA 02138}
\email{hguenther@cfa.harvard.edu}
\and

\author{C.~M.~Lisse}
\affil{Johns Hopkins University Applied Physics Laboratory, 11100 Johns Hopkins Road, Laurel, MD 20723}
\and
\author{J.~Robrade, J.~H.~M.~M.~Schmitt}
\affil{Hamburger Sternwarte, Universit\"{a}t Hamburg, Gojenbergsweg 112, 21029 Hamburg, Germany}

\begin{abstract}
A type stars are expected to be X-ray dark, yet weak emission has been detected from several objects in this class. We present new \emph{Chandra}/HRC-I observations of the A5~V star $\beta$~Pictoris. It is clearly detected with a flux of $(9\pm2)\times 10^{-4}$~counts~s$^{-1}$. In comparison with previous data this constrains the emission mechanism and we find that the most likely explanation is an optically thin, collisionally dominated, thermal emission component with a temperature around 1.1~MK. We interpret this component as a very cool and dim corona, with $\log L_X/L_{bol}=-8.2$ (0.2-2.0~keV). Thus, it seems that $\beta$~Pictoris shares more characteristics with cool stars than previously thought. 
\end{abstract}


\section{Introduction}
Stars across the main sequence (MS) produce X-ray emission by two fundamentally different mechanisms. Late-type stars have convective envelopes, thus they develop a solar-like dynamo which creates a magnetic field. In turn, they display magnetic activity in a corona. All late-type MS stars close enough to be detected in X-ray surveys are confirmed X-ray emitters \citep{2004A&A...417..651S}.  Similarly, probably all stars earlier than mid-B are X-ray emitters, a small number of non-detections among the more nearby objects likely being due to absorption \citep{1996A&AS..118..481B,1997A&A...322..167B}.
Early stars have supersonic, radiatively-driven winds. Instabilities in the winds shock-heat the gas to a few MK. Stars from mid-A to mid-B operate neither mechanism: Their winds are too weak to produce detectable X-ray emission and their atmospheres are radiatively dominated and do not drive a solar-like dynamo and thus are, at least in principle, X-ray dark \citep{1997A&A...318..215S}.

It is no contradiction that some of these systems are detected in the ROSAT All-Sky Survey (RASS), because they often have unresolved late-type companions. Due to the shorter lifetime of the A-type star the companion is still at an early stage of its evolution and thus X-ray bright. The RASS catalog contains 312 bright A-type stars, a detection rate of 10-15\% \citep{2007A&A...475..677S}. Most of those sources that have been re-observed at higher spatial resolution turn out to be multiple \citep{2000A&A...359..227H,2003A&A...407.1067S}, but the possibility remains that a fraction are bona-fide  X-ray emitting A stars.

We mention two classes as exceptions to the rule of X-ray dark mid-B to mid-A stars: (i) Ap/Bp stars such as IQ~Aur have strong magnetic fields and can funnel their winds to collide in the equatorial plane, forming shocks \citep{1997A&A...323..121B,2011A&A...531A..58R} and (ii) Herbig Ae/Be stars are young pre-main sequence stars of spectral type A or B with ages of only a few Myr. Herbig Ae/Be stars have significant mass accretion and sometimes strong jets. X-ray grating spectroscopy indicates both a coronal component from primordial fields or a turbulent dynamo and a soft X-ray component that is formed above the stellar surface, which has been interpreted as a jet base shock \citep{ABAur,HD163296}.

Recently, \object{Altair}, spectral type A7, has been observed with \emph{XMM-Newton} \citep{2009A&A...497..511R}. The star shows modest variability on the 30\% level, which is due to stellar rotation. The plasma temperature is 1-4~MK and the luminosity $L_X=1.4\times 10^{27}$~erg~s$^{-1}$. The activity level of $\log L_X/L_{bol}= -7.4$  is far below that of  saturated late-type stars which show $\log L_X/L_{bol}= -3$. In contrast, all single, earlier A stars remain undetected; the most prominent example is the A0V star \object{Vega}, where 29~ks of \emph{Chandra} ACIS and HRC observations place a 99.7\% upper limit as low as $2 \times 10^{25}$~ergs$^{-1}$ or $\log L_X/L_{bol}< -10$ (assuming a corona with $T = 1.5$~MK), one of the lowest limits ever obtained for an X-ray luminosity \citep{2006ApJ...636..426P,2008AJ....136.1810A}. Vega is thought to be up to 500~Myr old and it seems natural to expect higher luminosities from younger stars. We have recently observed the 8~Myr old early A-type star HR~4796A but failed to detect it down to a limit of $L_X=1.3\times 10^{27}$~erg~$^{-1}$ (Drake et al, submitted).

This article presents new \emph{Chandra}/HRC-I observations of the only known isolated mid A-type star with detected X-ray emission: the A5 star \object[beta Pic]{$\beta$~Pic} \citep{2005A&A...440..727H} at a distance of 19.4 pc. 
$\beta$~Pic is the eponymous member of the $\beta$~Pic
moving group (BPMG) of co-eval stars.  The age is somewhat uncertain and
ranges from 12 \citep{2001ApJ...562L..87Z} to 40~Myrs \citep{2010ApJ...723.1599M}. 
$\beta$~Pic harbors a bright debris disk \citep{1984Sci...226.1421S}, further characterized by many groups and telescopes, including Subaru \citep{2004ApJ...610L..49H} and Spitzer \citep{2007ApJ...666..466C}. The disk of $\beta$~Pic has been subject to intensive study, including IR-interferometry, which excludes any companion more massive than $50M_{Jup}$ in the inner few AU at at 90\% level \citep{2010A&A...520L...2A}. However, \citet{2009A&A...493L..21L} imaged a giant planet in the debris disk at around 8~AU from the primary, which can account for some of the disk substructure, and two other planets are probably located near the other dust belt concentrations \citep{2004ApJ...610L..49H}. The entire disk extent is some 1000~AU, and is dominated by icy dust in an extended Kuiper belt full of primitive comets and icy dwarf planets.
From this disk comets, or ``falling evaporating bodies (FEBs)''  appear to impact on the star on a daily basis \citep{1987A&A...185..267F}.

Observations with \emph{FUSE} in the FUV show strong and super-rotationally broadened lines of highly ionized species up to \ion{O}{6}. These lines consist of several components. Their emission mechanism is uncertain, but could be related to some sort of a chromosphere or cool corona a few stellar radii above the photosphere \citep{2001ApJ...557L..67D,2002A&A...390.1049B} or to accretion of mass from the debris disk. A further peculiarity is that  $\beta$~Pic shows $\delta$~Scuti type pulsations \citep{2003MNRAS.344.1250K}.


\section{Observations and data reduction}

We performed X-ray observations with \emph{Chandra}/HRC-I to confirm the \emph{XMM-Newton} detection of $\beta$~Pic and constrain the X-ray emission mechanism.
The HRC is a microchannel plate detector with excellent timing and spatial resolution, but only very limited energy resolution. The spatial size of the digitized pixels is 0.13175\arcsec{}. This is well matched to the intrinsic point-spread function, which can be described by a Gaussian with FWHM of 0.4\arcsec{} \citep{2000SPIE.4012...68M}. It also has excellent sensitivity to soft X-rays. Additional \emph{Chandra}/ACIS data were retrieved from the archive.
In contrast to the HRC, the ACIS CCDs have an intrinsic energy resolution. \emph{Chandra} data for the HRC-I and ACIS-I observations were reprocessed with CIAO 4.4 \citep{2006SPIE.6270E..60F} to extract event lists. The background lightcurves for both observations are flat confirming that no flaring of the ambient spacecraft proton impact rate was present.

Table~\ref{tab:obslog} contains details of all \emph{Chandra} and \emph{XMM-Newton} observations of $\beta$~Pic. Existing \emph{ROSAT} data is dominated by UV contamination but could have detected $\beta$~Pic if it were a few time brighter than in the \emph{XMM-Newton} observation \citep{2005A&A...440..727H}. We reprocessed the archival \emph{XMM-Newton} data with the current version 11.0.0 of the standard Science Analysis System (SAS) software \citep{2004ASPC..314..759G} with all standard selection criteria and refer the reader to \citet{2005A&A...440..727H} for further details of the \emph{XMM-Newton} observation.

\section{Source properties}

The \emph{XMM-Newton}/PN spectrum shows strong UV contamination due to the insufficient UV blocking used, but $\beta$~Pic can be detected in a very narrow energy filter around the \ion{O}{7} triplet in the MOS. Our analysis confirms the published results by \citet{2005A&A...440..727H}, who detect the source with a total count number of 17 (of which 6.1 are expected to be background photons) in both MOS detectors, a detection with a formal significance of 99.99\%. The flux of the \ion{O}{7} triplet is $3\times10^{25}$~erg~s$^{-1}$. 

In the ACIS-I observation there are 4 counts within a source extraction region of 1\arcsec{} radius centered on the optical position of $\beta$~Pic (05:47:17.08858 -51:03:59.2035). The individual photon energies are 280~eV, 300~eV, 720~eV and 11~keV, the last one is certainly a background event. From a large background region we estimate the background within the source region to be 1.5 counts, predominantly at high energies. The UV contamination of the ACIS data is negligible \citep[][with a 2001 update to account for an optical light leak for very red stars]{ACISCal}. 

If we scale the \ion{O}{7} flux in the 465-665~eV range from \citet{2005A&A...440..727H} to the shorter exposure time in the \emph{Chandra} observation and the lower effective area of the ACIS detector, we estimate that one photon from the \ion{O}{7} triplet should be recorded in 10~ks. $\beta$~Pic remains indeed undetected in an energy filter centered on \ion{O}{7} with 0 counts observed and an estimated background of 0.25 counts.

\begin{figure}
\plotone{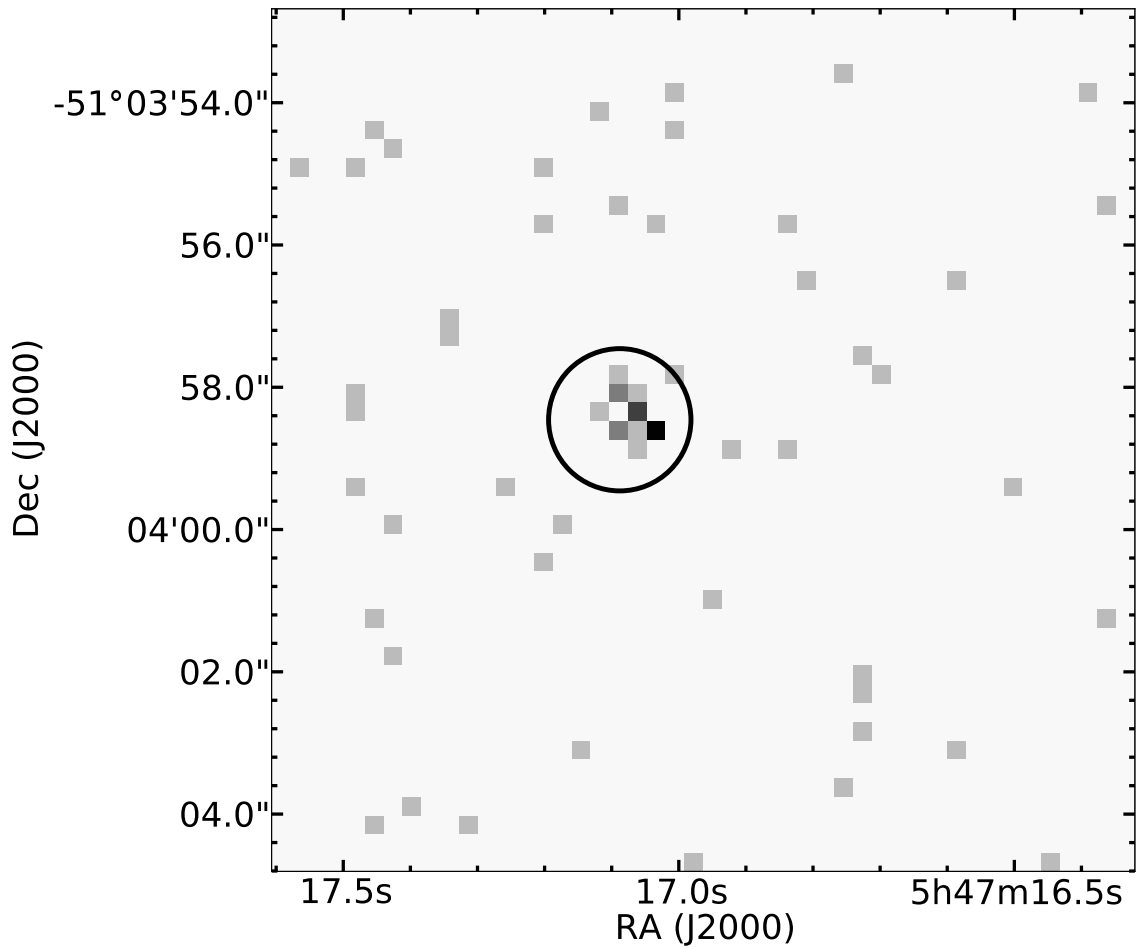}
\caption{HRC-I counts, binned by a factor 2. The shade indicates the number of counts per bin from light gray (1 count) to black (4 counts). The circle marks the extraction region centered on the optical position of $\beta$~Pic. \label{fig:detection}}
\end{figure}

Source detection on the HRC-I data is performed using the \texttt{celldetect} algorithm as implemented in CIAO. 
Our target $\beta$~Pic is detected with 17 counts in a circle of 1.0\arcsec{} radius (90\% encircled energy), while the background is estimated from a larger annulus around the source position (Figure~\ref{fig:detection}) to be $0.90\pm0.02$ counts. In addition to the uniform background, there is also a contribution caused by optical and UV light from $\beta$~Pic which leaks through the UV/Ion shield of the HRC-I. Scaling down the observed count rate of Vega\footnote{http://cxc.harvard.edu/contrib/juda/memos/uvis\_monitor/index.html} of 0.0005~cts~s$^{-1}$ to the $B$ and $V$ magnitude of $\beta$~Pic yields an estimated contribution of 0.5 counts over our exposure time. We also estimate the UV leak following \citet{2000SPIE.4012..467K} and \citet{Zombeck2000}, who developed a model for the out-of-band UV-optical effective area and folded them with UV and optical stellar spectra. This calculation gives 1 count over our exposure time, but we note that, again in comparison to Vega, this model is known to overestimate the light leak by a factor of 2-4.

The net source count rate is thus $(8\pm2)\times10^{-4}$~s$^{-1}$, which leads to $(9\pm2)\times10^{-4}$~counts~s$^{-1}$ considering that our extraction region contains 90\% of the PSF. The detected source position is RA=$86.82109\pm0.00003$, DEC=$-51.06622\pm0.00002$. The distance to the SIMBAD position of $\beta$~Pic at the time of observation (including proper motion) is 0.08\arcsec{}, which is well within the absolute \emph{Chandra} pointing accuracy.

While the error on the flux is large, the detection itself is significant, because counts are distributed according to Poisson statistics which deviates from the Gaussian approximation for low count numbers. The formal probability to observe 17 or more counts by chance, if only 1.5 counts are expected from the background and UV leak, is negligibly small.


Thus, we conclude that $\beta$~Pic is clearly detected in the HRC-I data. According to a KS test the photon arrival times are compatible (on the 86\% level) with a constant luminosity during the observation.

The low intrinsic energy resolution of the HRC-I and the low number of counts do not allow the calculation of any meaningful hardness ratios. The detector is sensitive from a few keV down to $\approx0.06$~keV, although the effective area drops from a peak of over 200~cm$^2$ at 1~keV to below 10~cm$^2$ for energies below 0.15~keV. 

\section{Discussion}
In all X-ray observations $\beta$~Pic either remained undetected or was seen only with very few photons. The number of photons observed in the \emph{XMM-Newton} and the \emph{Chandra}/ACIS observations are consistent with the \emph{ROSAT} limit, although $\beta$ Pic is not formally detected in ACIS. This shows that $\beta$~Pic was not significantly brighter in 1996 or 2002 than in 2004. If we assume that the emission is constant, we can estimate the count rate expected in the HRC-I due to the \ion{O}{7} flux seen in \emph{XMM-Newton} to be $5\times10^{-4}$~s$^{-1}$ according to WebPIMMS\footnote{http://heasarc.nasa.gov/Tools/w3pimms.html}. This is about half of the observed count rate. Since the upper limits on flux at other bands are relatively strict in the \emph{XMM-Newton} observations \citep{2005A&A...440..727H}, most of the remaining count rate must be caused by very soft emission in the 0.06-0.2~keV range where the MOS detectors are not sensitive. Because the effective area of the HRC-I depends strongly on the photon energy, a model is required to turn the count rate into an energy flux. Adopting a photon energy of 0.1~keV for all photons (monochromatic emission) we estimate the HRC-I flux to be $3\times10^{-14}$~erg~s$^{-1}$~cm$^{-2}$, i.e. $\log L_X = 27.1$. 

\subsection{A thermal plasma}
\begin{figure}
\plotone{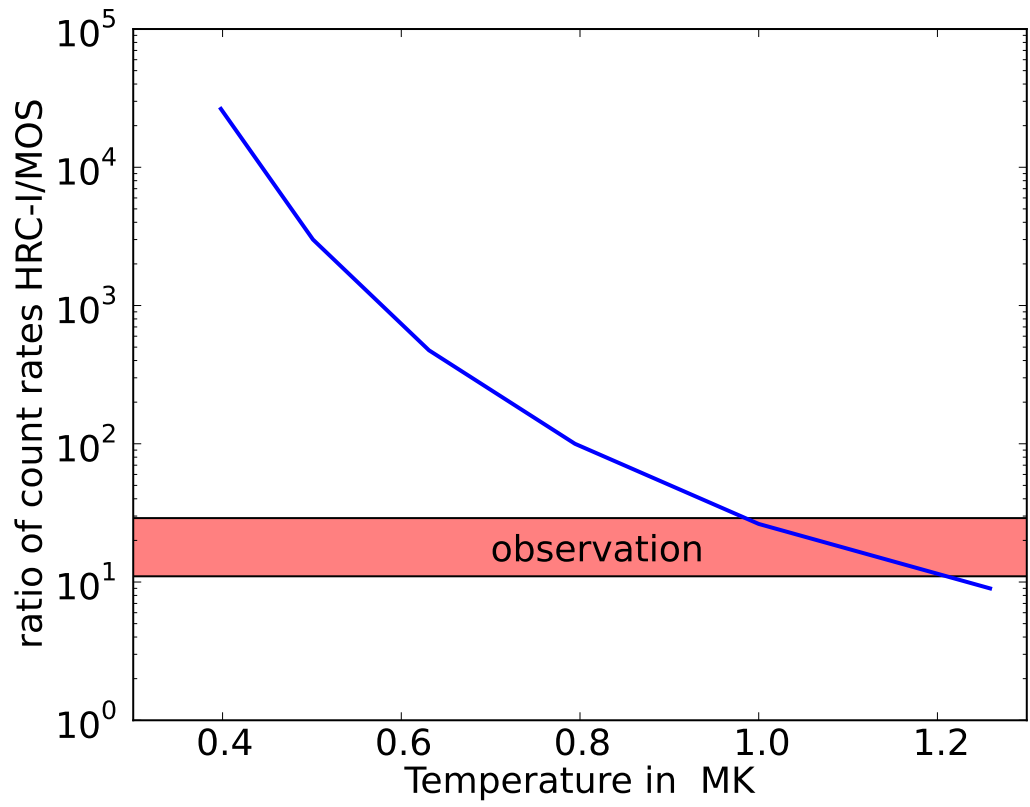}
\caption{Ratio of HRC-I counts (full band) and MOS counts (\ion{O}{7}) as a function of plasma temperature for an optically thin, collisionally excited thermal plasma with solar abundances. \label{fig:countrate}}
\end{figure}

The easiest explanation for the observed data is optically thin thermal emission. Fig.~\ref{fig:countrate} shows the ratio of HRC-I counts (full band) and MOS counts (\ion{O}{7}) as a function of plasma temperature for an optically thin, collisionally excited thermal plasma with solar abundances. Count rates for this figure were calculated using an APEC model \citep{2005AIPC..774..405B} and convolved with the respective effective areas using the WebPIMMS tool.
A plasma with a temperature of $1.1\times10^6$~K  
would explain the MOS and HRC-I data. This model gives $\log L_X = 26.5$ (0.2-2.0~keV) and $\log L_X = 27.5$ (0.06-5.~keV). In the following we use the energy band 0.2-2.0~keV, which allows for easy comparison to other observations. This leads to the following ratio of X-ray to bolometric luminosity: $\log L_X/L_{bol}=-8.2$, where we use $L_* = 5.0 \times 10^{34}$~erg~s$^{-1}$ \citep{Allen2000}.  Coronae of typical late-type stars with magnetic activity are generally much hotter than $\beta$~Pic,  and even the main-sequence A7 star Altair has a mean temperature twice as high and shows a small flare \citep{2009A&A...497..511R}. Altair is of later spectral type and the equatorial bulge is even cooler than the stellar $T_{eff}$ \citep{2007Sci...317..342M} and thus it will have a thin outer convective layer \citep{2009A&A...503..973L}. Altair's fast rotation could provide a magnetic field to drive magnetic activity on a similar level as on the quiescent sun. $\beta$~Pic rotates more slowly than Altair, but still has $v\sin i = 124\pm3$~km~s$^{-1}$ \citep{2003MNRAS.344.1250K} and might operate a similar mechanism. $\beta$~Pic is fainter than Altair which has $\log L_X = 27.1$ in the 0.2-2~keV band. Also, because $\beta$~Pic is younger there are alternative possibilities to explain a magnetic field, which could be primordial or formed by a dynamo which feeds of residual shear in the stellar atmosphere \citep{2002A&A...381..923S,1995MNRAS.272..528T}. Alternatively, the $\delta$~Scuti pulsations might supply energy to heat a chromosphere \citep{2003MNRAS.344.1250K} and possibly a cool and weak corona. In both cases the magnetically active regions can be distributed unevenly on the surface. Impacting FEBs might provide additional energy to excite waves which can heat a transition region and a corona.

The described thermal component can also account for some of the \ion{O}{6} emission seen by FUSE \citep{2001ApJ...557L..67D} but this fraction is strongly temperature dependent. 
Alternatively, \citet{2002A&A...390.1049B} present a model for the corona and the transition region that explains the observed UV lines very well. If the chromosphere covers a significant part of the stellar surface and rotates rigidly with the photosphere, then the large \ion{O}{6} line width is due to rotational broadening. Based on a comparison of the chromospheric activity on $\beta$~Pic and $\alpha$~Aql they predict an X-ray activity level on $\beta$~Pic of $\log L_X/L_{bol} = -7$, close to the value we observe.

In summary, the X-ray emission from $\beta$~Pic is fully consistent with a corona. The activity level is below that of Altair which is of slightly later type and much older. A convective dynamo that fades to earlier spectral types can explain both observations. In contrast, a shear dynamo scenario would be unique to $\beta$~Pic, since Altair is so old, that any initial shear would have decayed already  \citep{1995MNRAS.272..528T}.

\subsection{Accretion}
\citet{2005A&A...440..727H} suggested accretion of gas from the disk onto $\beta$~Pic itself as one possibility for the energy source of the observed UV and X-ray emission. While small mass accretion rates consistent with optical observations release sufficient power, the accretion streams would originate at the inner truncation radius of the disk and accelerate to the free fall velocity of a few hundred km~s$^{-1}$. The infalling matter should emit Doppler-shifted H$\alpha$ and other hydrogen lines with velocities up to the free-fall velocity, which is not observed \citep{2001ApJ...557L..67D}. To avoid this problem \citet{2005A&A...440..727H} propose a disk reaching down to the star, so that the accretion can proceed more gradually. This model predicts a boundary layer with lower velocities and thus lower temperatures around 0.3~MK. However, gas at this temperature would be much brighter in the HRC-I than observed (Fig.~\ref{fig:countrate}) and it should cause additional UV continuum emission, again contrary to observations \citep{2001ApJ...557L..67D}. Thus, an accretion scenario is unlikely to explain the observations.

\subsection{Charge-exchange}
Soft X-ray emission, typically dominated by a strong \ion{O}{7} triplet, has been observed for over 20 comets to date \citep[see reviews by][]{2004SSRv..113..271K,2007P&SS...55.1135B}. 
Ranging in luminosity from $5\times10^{13}$ to $5\times10^{15}$~ergs~s$^{-1}$ in the 300-1000 eV range, comets shine in the X-ray range as the neutral gas sublimating from their surface close to perihelion passage collides with the highly ionized solar wind and exchanges electrons. The electrons are transferred from the neutral gases into the n=4 to n=6 states of the wind ions, and then de-excite over the next few minutes, emitting characteristic X-rays from hydrogenic and heliogenic O, C, N, Ne, Fe, Si, and Mg. Charge exchange induced X-rays are also seen emanating from the heliosphere as a whole, produced when instreaming neutral ISM H and He atoms collide with the solar wind. The total luminosity for the latter process is on the order of $10^{23}$~erg~s$^{-1}$. Thus CXE could be producing the observed X-ray emission, if $10^{10}$ to $10^{12}$ solar system-like comets were extant in the disk; while the FEB detection suggests there may be many comets and KBOs in the disk, this seems like a huge number and would add up to as much as $1M_{Jup}$.

A second possibility is astrospheric emission detection from an interaction of the stellar wind and the inflowing ISM. A star with a strong wind would cause a large heliosphere, so that a charge-exchange can happen over a large region.

However, both these scenarios require a strong stellar wind with highly ionized ions. A type stars in general are expected to show only weak radiatively driven winds and on $\beta$~Pic \citet{1991ApJ...371L..27B} observed a slow wind with an outflow velocity of only 60~km~s$^{-1}$ and mass loss rates of $1.1\times10^{-14}\;M_{\sun}$~yr$^{-1}$. The stellar radiation field is insufficient to produce highly ionized ions like \ion{O}{7} in the wind and \citet{1993Icar..106...42B} showed that shocks between the stellar wind and the infalling comets are not strong enough to heat the gas to high ionization stages.

In summary, it is unlikely that highly ionized ions are present in the wind from $\beta$~Pic and even if there are charge exchange reactions with comets or the ISM they are still unable to explain the observed X-ray luminosity.


\section{Summary}
We have detected $\beta$~Pic in an \emph{Chandra}/HRC-I observation with a count rate of $(9\pm2)\times10^{-4}$~counts~s$^{-1}$ (corrected for the encircled fraction of the PSF). For a thermal model with $kT=0.1$~keV this corresponds to $\log L_X = 26.5$ in the 0.2-2.0~keV band. Ninety percent of this emission comes from outside of the \ion{O}{7} line. The emission can be explained by optically thin thermal emission of a cool corona with a temperature of 1.1~MK which sits on top of a solar-like chromosphere. Several dynamo scenarios could supply the energy for modest magnetic heating. We find that a dynamo in a weak convection zone (e.g. on the equatorial bulge) as proposed for Altair can explain all X-ray observations of A type stars consistently. Thus, the high energy emission of $\beta$~Pic might resemble our sun more closely than previously thought, despite the fact that $\beta$~Pic is surrounded by a massive debris disk and has a much earlier spectral type. With the exception of the accreting HAeBes, $\beta$~Pic is now the hottest star known with a solar-like corona.

\acknowledgements 
Support for this work was provided by the National Aeronautics and Space Administration through Chandra Award Number GO2-13015X issued by the Chandra X-ray Observatory Center, which is operated by the Smithsonian Astrophysical Observatory for and on behalf of the National Aeronautics Space Administration under contract NAS8-03060.

{\it Facilities:} \facility{CXO (HRC/ACIS)} \facility{XMM}


%
\bibliographystyle{../AAStex/astronat/apj/apj}
\bibliography{../articles}


\begin{deluxetable}{llrl}
\tablecaption{Log of observations\label{tab:obslog}}
\tablewidth{0pt}
\tablehead{
\colhead{Facility} & \colhead{Id} & \colhead{exposure} & \colhead{date}}
\startdata
Chandra/ACIS-I&\dataset[ADS/Sa.CXO#obs/02537]{2537} & 10 ks & 2002-10-09\\
XMM-Newton & 0044740601 & 72 ks & 2004-01-04 \\
Chandra/HRC-I &\dataset[ADS/Sa.CXO#obs/13626]{13626}& 19 ks & 2011-10-08\\
\enddata
\end{deluxetable}

\end{document}